\documentclass[sigconf]{acmart}

\setcopyright{rightsretained}
\copyrightyear{2026}
\acmConference[MSR '26]{23rd International Conference on Mining Software Repositories}{April 13--14,
  2026}{Rio de Janeiro, Brazil}

\usepackage{longtable}
\usepackage{booktabs}
\usepackage{array}
\usepackage{xspace}
\usepackage{tikz}
\usetikzlibrary{positioning, shapes.geometric}

\newcommand{\etc}{etc.\xspace}
\newcommand{\etal}{et al.\xspace}
\newcommand{\ie}{i.e.\xspace}
\newcommand{\eg}{e.g.\xspace}

\newcommand{\secref}[1]{Section~\ref{#1}\xspace}

\newcommand{\tabref}[1]{Table~\ref{#1}\xspace}

\usepackage{ifthen}                      
\usepackage[most]{tcolorbox}             
\newcommand{\importantstyle}{box}
\newcommand{\setBoxStyle}[1]{%
  \renewcommand{\importantstyle}{#1}}
\newenvironment{ResultBox}[2]{%
  \ifthenelse{\equal{\importantstyle}{box}}{%
    \begin{tcolorbox}[enhanced,
      title=#1,
      colback=#2!5,
      colframe=#2!75!black,
      boxrule=0.8pt,
      arc=4mm,outer arc=4mm,
      left=3mm,right=3mm,top=2mm,bottom=2mm]
  }{%
    \begin{tcolorbox}[enhanced,
      title=#1,
      borderline west={2pt}{0pt}{#2!75!black},
      colback=white,
      colframe=#2!50,
      sharp corners,
      boxrule=0.5pt,
      left=2mm,right=2mm,top=1mm,bottom=1mm]
  }%
}{%
  \end{tcolorbox}%
}
\newcounter{promisecounter}
\newenvironment{Promise}[1]{%
  \refstepcounter{promisecounter}%
  \begin{ResultBox}{Promise \thepromisecounter: #1}{blue}%
}{%
 
  \end{ResultBox}%
}

\newcounter{perilcounter}
\newenvironment{Peril}[1]{%
  \refstepcounter{perilcounter}%
  \begin{ResultBox}{Peril \theperilcounter: #1}{red}%
}{%
  \end{ResultBox}%
}

\newcommand{\basemitigation}[2]{
\par\textbf{#1}\\
\label{#2}
}

\newcommand{\mitigation}[2]{
\basemitigation{Mitigation for Peril~\theperilcounter: #1}{#2}
}

\begin{document}

\title[Mining Coding Agent Activity: Promises and Perils]{Promises, Perils, and (Timely) Heuristics\\ for Mining Coding Agent Activity}

\author{Romain Robbes}
\orcid{0000-0003-4569-6868}
\affiliation{%
  \institution{Univ. Bordeaux, CNRS, Bordeaux INP, LaBRI}
  \city{Bordeaux}
  \country{France}}
\email{romain.robbes@labri.fr}

\author{Théo Matricon}
\orcid{0000-0002-5043-3221}
\affiliation{%
  \institution{Univ Rennes, Inria, CNRS, IRISA}
  \city{Rennes}
  \country{France}}
\email{theo.matricon@inria.fr}

\author{Thomas Degueule}
\orcid{0000-0002-5961-7940}
\affiliation{%
  \institution{Univ. Bordeaux, CNRS, Bordeaux INP, LaBRI}
  \city{Bordeaux}
  \country{France}}
\email{thomas.degueule@labri.fr}

\author{Andre Hora}
\orcid{0000-0003-4900-1330}
\affiliation{%
  \institution{Department of Computer Science, UFMG}
  \city{Belo Horizonte}
  \country{Brazil}}
\email{andrehora@dcc.ufmg.br}

\author{Stefano Zacchiroli}
\orcid{0000-0002-4576-136X}
\affiliation{%
  \institution{LTCI, Télécom Paris, Institut Polytechnique de Paris}
  \city{Palaiseau}
  \country{France}}
\email{stefano.zacchiroli@telecom-paris.fr}

\renewcommand{\shortauthors}{Robbes et al.}

\begin{abstract}
In 2025, coding agents have seen a very rapid adoption. Coding agents leverage Large Language Models (LLMs) in ways that are markedly different from LLM-based code completion, making their study critical. Moreover, unlike LLM-based completion, coding agents leave visible traces in software repositories, enabling the use of MSR techniques to study their impact on SE practices. This paper documents the promises, perils, and heuristics that we have gathered from studying coding agent activity on GitHub.
\end{abstract}

\maketitle

\section{Introduction}
Large Language Models (LLMs) have had a major impact on the practice of Software Engineering~\cite{fan2023large}.
One of the first LLMs to be used in practice was Codex~\cite{chen2021evaluating}, which was the model powering the initial version of GitHub's coding assistant, Copilot, starting in 2021.
Codex and Copilot immediately triggered a flurry of empirical studies about both the capacities and limitations of the underlying model, and how the tool impacted software development (\secref{sec:related} details these studies).
Unfortunately, few of these studies used Mining Software Repositories (MSR) techniques; they relied instead on controlled experiments~\cite{peng2023impact}, observation studies~\cite{barke2023grounded}, or surveys~\cite{liang2024large}.
While valuable, the lack of large-scale studies of the phenomenon limits the generalizability of the findings~\cite{stol2018abc}.
The dearth of MSR studies is for good reason: since coding assistant are used interactively in a code editor, it is essentially impossible to know whether they were used to author code in a repository, leaving studies with very rough time-based heuristics as the only option~\cite{hardinggitclear2025}, or studies of other LLMs uses~\cite{tufano2024unveiling}.

In 2025, a second generation of LLM-powered tools emerged and is seeing rapid adoption: coding agents.
(See \secref{sec:agents} for a detailed background on coding agents.)
If coding assistants are used interactively to complete lines or blocks of code, the scope of coding agents is drastically broader.
The key characteristic of coding agents is \emph{autonomy}. In ideal scenarios, a developer delegates an entire task to a coding agent, which performs a sequence of actions in the code base to accomplish the task, producing a code change that the developer can review, as a human developer would. 
The arrival of reasoning models in late 2024, and the focus of AI industrial labs to improve tool-calling capabilities allowed agents to quickly transition from promising academic work in 2024~\cite{zhang2024autocoderover, bouzenia2025repairagent, xia2024agentless} all the way to an established tool category in 2025, where all major AI labs, IDE vendors, and multiple startups propose products.
Adoption has quickly followed: our study of the topic, based on the heuristics presented here, found that between 15 and 20\% of GitHub projects adopted coding agents to some extent, a very large number for products that have been for the vast majority released this year~\cite{agentadoptionpaper}. 
Interest and adoption are very high because of the potential of coding agents.
While the scope of the tasks for which they can be used in full autonomy is limited today, improvements in model capabilities could change the situation.
If this promise materializes, this could represent a greater change to the software engineering practice than the first generation of LLM tools, and the arrival of CASE tools many decades ago.

Given the potential impact of coding agents, studying how they are used in practice is extremely important.
MSR studies have a major role to play here, providing an alternative viewpoint to controlled experiments~\cite{becker2025measuring} or observational studies~\cite{kumar2025sharp}.
It turns out that coding agents, by their very nature, leave much more traces in software repositories than code completion LLMs did, making their study via MSR techniques possible.
\secref{sec:heuristics} documents the heuristics we were able to identify in order to study this phenomenon, constituting the \emph{first contribution} of this paper. 

While coding agents leave traces in repositories that we can study, these come with important caveats: based on our experience studying agents, we have identified several of these, including that the observed data is partial, comes from multiple, heterogeneous agents, and is rapidly changing. On the other hand, there is ample space for future work studying the impact of coding agents, how they are used, how changes to the underlying LLMs affect them, and potentially changes to SE practices. In addition, as coding agent adoption is high, detecting and excluding coding agent activity will be necessary for MSR studies focusing on humans. We expand on the promises, perils and mitigations we formulated in \secref{sec:promises-perils}, which constitute our \emph{second contribution}.

Finally, as one of the perils we identify is \hyperref[per:velocity]{the rapid rate of change of coding agents}, no paper will remain the definitive resource for studying coding agent traces in software repositories for long.
While we expect the promises and perils to remain quite stable over time, heuristics to identify agent activity will change often.
This is why we call for the community to join us in documenting heuristics in a community-maintained repository, which we have bootstrapped \cite{agentminingrepo}.
The repository also contains sample datasets of traces found for specific coding agents, in order for researchers to investigate what coding agent traces look like and help them plan their studies before delving in massive amounts of real-world data.
This heuristic and data repository is our \emph{third contribution}; \secref{sec:repository} details it.
We discuss the limitations and implications of this work in \secref{sec:discussion}, before concluding with \secref{sec:conclusion}.



\section{Related work}
\label{sec:related}
Generative AI for SE and Coding assistants such as Github Copilot have been extensively studied, using a variety of techniques; we cite but a few, with a more specific focus on MSR studies.

\subsection{Studies of coding assistants}

\paragraph{Controlled experiments} Early studies of developer productivity showed mixed results. Peng \etal~\cite{peng2023impact} found 95 programmers completed HTTP servers 55\% faster using Copilot (71 vs 161 minutes). However, Imai~\cite{imai2022github} found higher productivity but lower code quality with Copilot versus pair programming, while Vaithilingam \etal~\cite{vaithilingam2022expectation} found participants failed tasks more often with Copilot than Intellisense due to incorrect code, despite preferring Copilot. Similarly, studies focused on code security yielded contrasting results. Sandoval \etal~\cite{sandoval2023lost} and Asare \etal~\cite{asare2023user} found little or no differences in terms of security aspects, while Perry \etal~\cite{perry2023users} found AI-assisted code was less secure across 6 tasks.

\paragraph{Qualitative studies} Barke \etal~\cite{barke2023grounded} identified two usage modes: in \emph{acceleration}, developers prefer small, quickly-validated suggestions; in \emph{exploration}, they prefer larger suggestions as starting points when working with unfamiliar APIs.
Mozannar \etal~\cite{mozannar2022reading} observed 21 programmers; they spent 22\% of time evaluating suggestions versus 14\% writing code, suggesting acceptance rates underestimate cognitive overhead.
Wang \etal~\cite{wang2023investigating} interviewed 17 developers who emphasized understanding Copilot's limitations and expressed concerns about negative impacts on learning.

\subsection{MSR studies of generative AI}

\paragraph{Telemetry} Telemetry-based approaches are possible in scenarios where the deployment can be carefully controlled, such as in companies developing the tools themselves. Murali \etal~\cite{murali2023codecompose} reported Meta's CodeCompose achieved 22\% acceptance rate and generated 8\% of company-wide code. Ziegler \etal~\cite{ziegler2022productivity} found acceptance rate was the best predictor of perceived productivity among 2,000 Copilot users. Izadi \etal~\cite{izadi2024language} developed a bespoke IDE extension and collected interaction data of ~1200 users, and analyzed model failures qualitatively.

\paragraph{Usage in software repositories} Tufano \etal~\cite{tufano2024unveiling} searched for explicit usage of ChatGPT in GitHub commits, pull requests, and issues. They analyzed ChatGPT usage traces in 467 GitHub instances, developing a taxonomy of 45 software engineering tasks including feature implementation, documentation, software quality, and development processes. Some usage patterns were unexpected, such as assistance in motivating proposed changes.

Xiao \etal~\cite{xiao2025self_admitted_ai} also look for usage of ChatGPT or Copilot when explicitly written by developers in software artifacts. They analyze more than 1,200 Generative AI usages in a qualitative study, focusing on the type of tasks they were used in; in addition they also analyzed instances of guidelines to regulate the usage of Generative AI in several projects, and also analyzed churn, finding overall no significant change in churn post-adoption 

This result is in contrast to the white papers authored by GitClear~\cite{hardinggitclear2024, hardinggitclear2025}, where they compared code churn before and after the introduction of Copilot, taken as a marker for the first usages of Generative AI in software repositories. The studies find a higher rate of churn as time passes and (presumably) generative AI adoption increases, and a higher incidence of code duplication in the repositories they analyze. However, in the absence of specific markers to detect the usage of Generative AI it is more difficult to untangle the effects of Generative AI from other factors. 

\paragraph{MSR mining challenges} Xiao \etal~\cite{xiao2024devgpt} curated a dataset of shared ChatGPT conversations with ChatGPT on software engineering topics. This dataset was used for the 2024 challenge, in which a variety of aspects were analyzed by 18 challenge submissions. Concurrent to this work, Li \etal~\cite{li2025aidev} curated a dataset of agent pull requests, to be used for the 2026 challenge.


\subsection{Studies of coding agents}

Given their recency, studies of coding agents are still few and far-between. Given the small amount of studies and their mixed results, we see this as an ideal ground for MSR studies to complement this emerging body of evidence. 

Becker et al.~\cite{becker2025measuring} performed a controlled experiment of software developers using Cursor. Developers believed that using Cursor reduced completion time of tasks by 20\%, but it actually increased task completion time by 19\%. Many potential factors could explain this slowdown. Notable ones include less than 44\% of code generations accepted, 9\% of the time spent cleaning AI code. Importantly, the agent used had limited capabilities, and was used in interactive rather than autonomous mode.

Kumar et al.~\cite{kumar2025sharp} performed an observation study of 19 developers using Cursor. There were two categories of participants: those who tried to delegate their entire task to the agent, and those that collaborated with the agent to decompose the task in subtasks, and have the agent solve each subtask. Participants provided the agent several kinds of information, notably contextual information from the task description and the environment, and expert information based on prior repository knowledge. Indeed, the main barrier to effective agent use was when it lacked this tacit knowledge.

Bouzenia and Pradel~\cite{bouzenia2025understanding} analyzed 120 agent interaction logs as they were attempting to solve bugs from the SWE-bench benchmark. While these traces come from less established agents, and are not the result of genuine human interactions but rather byproducts of a benchmark, they provide useful insights, highlighting the differences in behaviour between the agents in terms of the length of their interaction traces, the type of interactions in them, and the differences between succesful and failing attempts.



\section{Coding agents}
\label{sec:agents}


While there are multiple definitions of an agent, we use the following one: an agent is an LLM executing in a loop in order to fulfill a goal, and that is provided access to tools in order to access its environments. The agent loop is then conceptually simple: 
\begin{enumerate}
\item query the LLM with a sequence of interactions (initially, just the task description);
\item parse the response for the presence of tool usages and end of task;
\item if tool calls are present: optionally ask the user for permission before executing sensitive tools;
\item if task has not ended: add the response to the sequence of interactions, replacing tool calls with their results.
\end{enumerate}
The loop executes until the LLM determines it has completed the task, in which case its response indicates it, ending the conversation.

The first component contributing to the agent is the LLM itself. Modern LLMs have improved significantly on important capabilities necessary for agentic use cases. In particular, the capability to reliably generate structured outputs is crucial for reliable tool calling, and the abilities of reasoning models to generate and exploit long reasoning traces are critical to solve longer tasks.

The agent's harness is the second component of an agent. The harness is responsible for providing the model with the list of available tools, and detecting when the model generates a tool call (using a structured format, such as JSON, containing \eg the name of the tool and its possible parameters). 

\paragraph{Tool use} Access to tools is a key differentiator with coding assistants. While coding assistants are provided a context for their completion, it is computed beforehand by the IDE. In contrast, tool use enables much more autonomous workflows. For example:
\begin{itemize}
\item using search tools and file reading tools, an agent can explore the code base and dynamically find the right context for its task, instead of relying on the IDE's context;
\item using file writing tools, a compiler, and executing test cases, the agent can iterate on its solution until it is correct, leveraging error messages (if any) to improve it;
\item if the agent has determined that it has completed the task, it may commit changes or author a pull request, summarizing the changes in the process.
\end{itemize}

Tools range from the simple to very complex: some agents use little more than shell access (the agent writes shell commands to support a variety of tools, e.g., calling the compiler, running tests, using git), up to using \eg the model context protocol to interface with complex systems such as controlling a web browser, querying a database, or using a ticket repository. 

With these capabilities coding agents have been employed to perform a variety of tasks, ranging from updating dependencies, fixing simple bugs, or writing documentation, to implementing larger scoped features and their tests, refactor code, or port software from one programming language to another. 

\paragraph{Autonomy and oversight} Agents can have varying degrees of autonomy, depending on the sensitivity of the task, the degree of oversight a developer wants to invest, the environment setup, and the individual workflow of each agent. Depending on their prompt, agent may ask developers for feedback on the task (e.g., write a plan to implement a feature and submit it for feedback). Moreover, some agents allow human oversight by interrupting the agent's activity and add additional guidance before resuming.

Agents may require human oversight when they are using tools; each tool can have individual permissions, ranging from allowing all uses of a given tool, allowing the use on a case-by-case basis (with human approval), to systematically forbidding a tool. A key advantage of allowing tool use without human intervention is that it makes the agent fully autonomous, which frees up the developer to work on other tasks, rather than closely steering the agent.  

\paragraph{Uses} There are a breadth of uses for coding agents, ranging from ``vibe coding''~\cite{Wikipedia2024VibeCoding, fawz2025vibe} with little to no supervision, up to serious practice with much more oversight~\cite{kumar2025sharp}. Coding agents already have the potential to help developers being more productive (although for early coding agents, the evidence is mixed~\cite{becker2025measuring}), and to feel less alienated in their work (e.g., by spending less time on menial tasks). Uses may evolve over time: in particular, technical solutions (e.g., development in a container~\cite{dagger_container_use, claude_code_devcontainer}) make full autonomy less risky as the impact of the agent is controlled. Some advocate to leverage the agent's autonomy to the maximum, such as the use of parallel agents working on multiple tasks at once.


\section{Traces and heuristics}
\label{sec:heuristics}

For coding agent studies~\cite{agentadoptionpaper}, we have gathered a set of heuristics that detect their presence in software repositories. We focus on GitHub, the most common coding platform; furthermore, due to its widespread use, some agents leverage some specificities of GitHub (e.g., activity in pull requests). We derive our heuristics from an extensive manual investigation of known agents, involving checking their documentation for mentions of specific artifacts, and analysis of repositories identified as using agents for visible traces. This leads to a tentative list, which we validate with targeted GitHub searches and manual checks to select heuristics that return enough results to be useful, and do not have too many false positives. 


\makeatletter
\newcommand{\heuristiclink}[3]{%
  \leavevmode
  \strut
  \ifnum#3<250
    \hbox{}
  \else\ifnum#3<1000
    {\color{gray}\href{#2}{#1}}%
    \@ifnextchar\heuristiclink{, }{}%
  \else\ifnum#3<5000
    \href{#2}{#1}%
    \@ifnextchar\heuristiclink{, }{}%
  \else\ifnum#3<25000
    \textit{\href{#2}{#1}}%
    \@ifnextchar\heuristiclink{, }{}%
  \else\ifnum#3<100000
    \textbf{\href{#2}{#1}}%
    \@ifnextchar\heuristiclink{, }{}%
  \else\ifnum#3<500000
    {\color{orange}\textbf{\href{#2}{#1}}}%
    \@ifnextchar\heuristiclink{, }{}%
  \else
    {\color{red}\textbf{\href{#2}{#1}}}%
    \@ifnextchar\heuristiclink{, }{}%
  \fi\fi\fi\fi\fi\fi
  \strut
}
\makeatother

\newcolumntype{P}[1]{>{\raggedright\arraybackslash\setlength{\parskip}{0pt}\setlength{\parindent}{0pt}}p{#1}}

\begin{table*}[htbp]
\caption{Agent detection heuristics and approximate GitHub match counts on 20/10/25 (click to browse results; file queries require GitHub login). The table includes agents with more than 1000 matches, and heuristics with more than 250 matches.}
\centering
\scriptsize
\begin{tabular}{P{2cm} P{5cm} P{4cm} P{1.8cm} P{1.8cm} P{1cm}}
\hline
\textbf{Tool} & \textbf{File} & \textbf{Commit co-author or author} & \textbf{Branches} & \textbf{Labels} & \textbf{Total} \\
\hline
\href{https://agents.md}{Generic} & \heuristiclink{AGENTS.md (37.5K)}{https://github.com/search?q=path\%3A/\%28\%3F\%3A\%5E\%7C\%5C/\%29\%28AGENTS\%5C.md\%29\%24/\&type=code}{37500} & - & - & \heuristiclink{ai-generated (1.9K)}{https://github.com/search?q=label\%3Aai-generated\%20type\%3Apr\&type=pullrequests}{1900} & 39.4K \\
\hline
\href{https://aider.chat/}{Aider} & \heuristiclink{.aider.conf.yml (422)}{https://github.com/search?q=path\%3A/\%28\%3F\%3A\%5E\%7C\%5C/\%29\%28\%5C.aider\%5C.conf\%5C.yml\%29\%24/\&type=code}{422} & \heuristiclink{aider (44.7K)}{https://github.com/search?q=Co-authored-by\%3A\%22aider\%22\&type=commits}{44700} \heuristiclink{aider@aider.chat (40.8K)}{https://github.com/search?q=Co-authored-by\%3A\%22aider\%40aider.chat\%22\&type=commits}{40800} & - & - & 85.9K \\
\hline
\href{https://sourcegraph.com/amp}{Amp} & \heuristiclink{AGENT.md (5.9K)}{https://github.com/search?q=path\%3A/\%28\%3F\%3A\%5E\%7C\%5C/\%29\%28AGENT\%5C.md\%29\%24/\&type=code}{5900} & \heuristiclink{amp@ampcode.com (5.9K)}{https://github.com/search?q=Co-authored-by\%3A\%22amp\%40ampcode.com\%22\&type=commits}{5900} & - & \heuristiclink{amp (260)}{https://github.com/search?q=label\%3Aamp\%20type\%3Apr\&type=pullrequests}{260} & 12.0K \\
\hline
\href{https://www.augmentcode.com/}{Augment Code} & \heuristiclink{.augment/ (1.4K)}{https://github.com/search?q=path\%3A/\%28\%3F\%3A\%5E\%7C\%5C/\%29\%28\%5C.augment\%5C/\%29/\&type=code}{1400} & - & - & - & 1.4K \\
\hline
\href{https://chatgpt.com/}{ChatGPT} & - & \heuristiclink{ChatGPT (23.3K)}{https://github.com/search?q=Co-authored-by\%3A\%22ChatGPT\%22\&type=commits}{23300} & - & - & 23.3K \\
\hline
\href{https://www.claude.com/product/claude-code}{Claude Code} & \heuristiclink{CLAUDE.md (110.0K)}{https://github.com/search?q=path\%3A/\%28\%3F\%3A\%5E\%7C\%5C/\%29\%28CLAUDE\%5C.md\%29\%24/\&type=code}{110000} \heuristiclink{.claude/ (141.0K)}{https://github.com/search?q=path\%3A/\%28\%3F\%3A\%5E\%7C\%5C/\%29\%28\%5C.claude\%5C/\%29/\&type=code}{141000} \heuristiclink{.claudeignore (290)}{https://github.com/search?q=path\%3A/\%28\%3F\%3A\%5E\%7C\%5C/\%29\%28\%5C.claudeignore\%29\%24/\&type=code}{290} \heuristiclink{.github/workflows/claude (5.4K)}{https://github.com/search?q=path\%3A/\%28\%3F\%3A\%5E\%7C\%5C/\%29\%28\%5C.github\%5C/workflows\%5C/claude\%29/\&type=code}{5400} & \heuristiclink{Claude (author, 38.2K)}{https://github.com/search?q=author\%3A\%22Claude\%22\&type=commits}{38200} \heuristiclink{noreply@anthropic.com (2.7M)}{https://github.com/search?q=Co-authored-by\%3A\%22noreply\%40anthropic.com\%22\&type=commits}{2700000} \heuristiclink{claude@anthropic.com (2.3K)}{https://github.com/search?q=Co-authored-by\%3A\%22claude\%40anthropic.com\%22\&type=commits}{2300} \heuristiclink{assistant@anthropic.com (314)}{https://github.com/search?q=Co-authored-by\%3A\%22assistant\%40anthropic.com\%22\&type=commits}{314} & \heuristiclink{claude/ (20.7K)}{https://github.com/search?q=head\%3Aclaude/\%20type\%3Apr\&type=pullrequests}{20700} & - & 3.0M \\
\hline
\href{https://cline.bot/}{Cline} & \heuristiclink{.clinerules (1.7K)}{https://github.com/search?q=path\%3A/\%28\%3F\%3A\%5E\%7C\%5C/\%29\%28\%5C.clinerules\%29\%24/\&type=code}{1700} \heuristiclink{.cline/ (558)}{https://github.com/search?q=path\%3A/\%28\%3F\%3A\%5E\%7C\%5C/\%29\%28\%5C.cline\%5C/\%29/\&type=code}{558} \heuristiclink{memory-bank/ (23.0K)}{https://github.com/search?q=path\%3A/\%28\%3F\%3A\%5E\%7C\%5C/\%29\%28memory\%5C-bank\%5C/\%29/\&type=code}{23000} \heuristiclink{memory\_bank/ (1.6K)}{https://github.com/search?q=path\%3A/\%28\%3F\%3A\%5E\%7C\%5C/\%29\%28memory\_bank\%5C/\%29/\&type=code}{1600} & \heuristiclink{cline (11.0K)}{https://github.com/search?q=Co-authored-by\%3A\%22cline\%22\&type=commits}{11000} \heuristiclink{cline@example.com (3.3K)}{https://github.com/search?q=Co-authored-by\%3A\%22cline\%40example.com\%22\&type=commits}{3300} & - & - & 41.1K \\
\hline
\href{https://codegen.com/}{Codegen} & - & \heuristiclink{codegen-sh (2.8K)}{https://github.com/search?q=Co-authored-by\%3A\%22codegen-sh\%22\&type=commits}{2800} & \heuristiclink{codegen-bot/ (6.1K)}{https://github.com/search?q=head\%3Acodegen-bot/\%20type\%3Apr\&type=pullrequests}{6100} & - & 8.9K \\
\hline
\href{https://coderabbit.ai/}{Coderabbit} & - & \heuristiclink{coderabbit (19.6K)}{https://github.com/search?q=Co-authored-by\%3A\%22coderabbit\%22\&type=commits}{19600} & - & - & 19.6K \\
\hline
\href{https://openai.com/codex}{Codex} & \heuristiclink{.codex/ (3.3K)}{https://github.com/search?q=path\%3A/\%28\%3F\%3A\%5E\%7C\%5C/\%29\%28\%5C.codex\%5C/\%29/\&type=code}{3300} & \heuristiclink{codex@openai.com (305)}{https://github.com/search?q=Co-authored-by\%3A\%22codex\%40openai.com\%22\&type=commits}{305} & \heuristiclink{codex/ (2.1M)}{https://github.com/search?q=head\%3Acodex/\%20type\%3Apr\&type=pullrequests}{2100000} & \heuristiclink{codex (2.3M)}{https://github.com/search?q=label\%3Acodex\%20type\%3Apr\&type=pullrequests}{2300000} & 4.4M \\
\hline
\href{https://www.continue.dev/}{Continue} & \heuristiclink{.continue/ (1.0K)}{https://github.com/search?q=path\%3A/\%28\%3F\%3A\%5E\%7C\%5C/\%29\%28\%5C.continue\%5C/\%29/\&type=code}{1000} & - & - & - & 1.0K \\
\hline
\href{https://github.com/features/copilot}{Copilot} & \heuristiclink{copilot-instructions.md (45.3K)}{https://github.com/search?q=path\%3A/\%28\%3F\%3A\%5E\%7C\%5C/\%29\%28copilot\%5C-instructions\%5C.md\%29\%24/\&type=code}{45300} \heuristiclink{copilot\_instructions.md (382)}{https://github.com/search?q=path\%3A/\%28\%3F\%3A\%5E\%7C\%5C/\%29\%28copilot\_instructions\%5C.md\%29\%24/\&type=code}{382} \heuristiclink{copilot-instructions/ (273)}{https://github.com/search?q=path\%3A/\%28\%3F\%3A\%5E\%7C\%5C/\%29\%28copilot\%5C-instructions\%5C/\%29/\&type=code}{273} \heuristiclink{.github/instructions/ (13.3K)}{https://github.com/search?q=path\%3A/\%28\%3F\%3A\%5E\%7C\%5C/\%29\%28\%5C.github\%5C/instructions\%5C/\%29/\&type=code}{13300} \heuristiclink{.copilotignore (382)}{https://github.com/search?q=path\%3A/\%28\%3F\%3A\%5E\%7C\%5C/\%29\%28\%5C.copilotignore\%29\%24/\&type=code}{382} \heuristiclink{.copilot/ (1.3K)}{https://github.com/search?q=path\%3A/\%28\%3F\%3A\%5E\%7C\%5C/\%29\%28\%5C.copilot\%5C/\%29/\&type=code}{1300} \heuristiclink{.github/workflows/copilot/ (1.6K)}{https://github.com/search?q=path\%3A/\%28\%3F\%3A\%5E\%7C\%5C/\%29\%28\%5C.github\%5C/workflows\%5C/copilot\%29/\&type=code}{1600} & \heuristiclink{Copilot (190.0K)}{https://github.com/search?q=Co-authored-by\%3A\%22Copilot\%22\&type=commits}{190000} \heuristiclink{copilot-swe-agent (35.5K)}{https://github.com/search?q=Co-authored-by\%3A\%22copilot-swe-agent\%22\&type=commits}{35500} & \heuristiclink{copilot/ (338.0K)}{https://github.com/search?q=head\%3Acopilot/\%20type\%3Apr\&type=pullrequests}{338000} & - & 626.0K \\
\hline
\href{https://github.com/charmbracelet/crush}{Crush} & \heuristiclink{CRUSH.md (382)}{https://github.com/search?q=path\%3A/\%28\%3F\%3A\%5E\%7C\%5C/\%29\%28CRUSH\%5C.md\%29\%24/\&type=code}{382} & \heuristiclink{crush@charm.land (1.3K)}{https://github.com/search?q=Co-authored-by\%3A\%22crush\%40charm.land\%22\&type=commits}{1300} & - & - & 1.6K \\
\hline
\href{https://cursor.com/}{Cursor} & \heuristiclink{.cursor/ (98.3K)}{https://github.com/search?q=path\%3A/\%28\%3F\%3A\%5E\%7C\%5C/\%29\%28\%5C.cursor\%5C/\%29/\&type=code}{98300} \heuristiclink{.cursorrules (16.6K)}{https://github.com/search?q=path\%3A/\%28\%3F\%3A\%5E\%7C\%5C/\%29\%28\%5C.cursorrules\%29\%24/\&type=code}{16600} \heuristiclink{CURSOR.md (1.3K)}{https://github.com/search?q=path\%3A/\%28\%3F\%3A\%5E\%7C\%5C/\%29\%28CURSOR\%5C.md\%29\%24/\&type=code}{1300} & \heuristiclink{cursor (206.0K)}{https://github.com/search?q=Co-authored-by\%3A\%22cursor\%22\&type=commits}{206000} \heuristiclink{cursoragent@cursor.com (40.7K)}{https://github.com/search?q=Co-authored-by\%3A\%22cursoragent\%40cursor.com\%22\&type=commits}{40700} & \heuristiclink{cursor/ (209.0K)}{https://github.com/search?q=head\%3Acursor/\%20type\%3Apr\&type=pullrequests}{209000} & - & 571.9K \\
\hline
\href{https://deepsource.com/}{Deepsource} & - & \heuristiclink{deepsource-autofix (10.2K)}{https://github.com/search?q=Co-authored-by\%3A\%22deepsource-autofix\%22\&type=commits}{10200} & - & - & 10.2K \\
\hline
\href{https://devin.ai/}{Devin} & - & \heuristiclink{devin-ai-integration (14.8K)}{https://github.com/search?q=Co-authored-by\%3A\%22devin-ai-integration\%22\&type=commits}{14800} & \heuristiclink{devin/ (49.8K)}{https://github.com/search?q=head\%3Adevin/\%20type\%3Apr\&type=pullrequests}{49800} & - & 64.6K \\
\hline
\href{https://gemini.google.com/}{Gemini} & \heuristiclink{GEMINI.md (8.4K)}{https://github.com/search?q=path\%3A/\%28\%3F\%3A\%5E\%7C\%5C/\%29\%28GEMINI\%5C.md\%29\%24/\&type=code}{8400} \heuristiclink{.gemini/ (4.8K)}{https://github.com/search?q=path\%3A/\%28\%3F\%3A\%5E\%7C\%5C/\%29\%28\%5C.gemini\%5C/\%29/\&type=code}{4800} & \heuristiclink{gemini-code-assist (22.5K)}{https://github.com/search?q=Co-authored-by\%3A\%22gemini-code-assist\%22\&type=commits}{22500} \heuristiclink{gemini-cli (3.5K)}{https://github.com/search?q=Co-authored-by\%3A\%22gemini-cli\%22\&type=commits}{3500} \heuristiclink{Gemini 2.5 Pro (11.2K)}{https://github.com/search?q=Co-authored-by\%3A\%22Gemini\%202.5\%20Pro\%22\&type=commits}{11200} \heuristiclink{Gemini 2.5 Flash (4.2K)}{https://github.com/search?q=Co-authored-by\%3A\%22Gemini\%202.5\%20Flash\%22\&type=commits}{4200} & - & - & 54.6K \\
\hline
\href{https://gru.ai/}{Gru} & - & \heuristiclink{gru-agent (2.4K)}{https://github.com/search?q=Co-authored-by\%3A\%22gru-agent\%22\&type=commits}{2400} & \heuristiclink{gru/ (1.3K)}{https://github.com/search?q=head\%3Agru/\%20type\%3Apr\&type=pullrequests}{1300} & - & 3.7K \\
\hline
\href{https://jules.google/}{Jules} & - & \heuristiclink{google-labs-jules (10.1K)}{https://github.com/search?q=Co-authored-by\%3A\%22google-labs-jules\%22\&type=commits}{10100} & \heuristiclink{jules/ (1.5K)}{https://github.com/search?q=head\%3Ajules/\%20type\%3Apr\&type=pullrequests}{1500} & - & 11.6K \\
\hline
\href{https://www.jetbrains.com/junie/}{Junie} & \heuristiclink{.junie/ (2.0K)}{https://github.com/search?q=path\%3A/\%28\%3F\%3A\%5E\%7C\%5C/\%29\%28\%5C.junie\%5C/\%29/\&type=code}{2000} & - & - & - & 2.0K \\
\hline
\href{https://kilocode.ai/}{Kilo Code} & \heuristiclink{.kilocode/ (1.8K)}{https://github.com/search?q=path\%3A/\%28\%3F\%3A\%5E\%7C\%5C/\%29\%28\%5C.kilocode\%5C/\%29/\&type=code}{1800} & \heuristiclink{Kilo Code (474)}{https://github.com/search?q=Co-authored-by\%3A\%22Kilo\%20Code\%22\&type=commits}{474} & - & - & 2.3K \\
\hline
\href{https://kiro.dev/}{Kiro} & \heuristiclink{.kiro/ (42.2K)}{https://github.com/search?q=path\%3A/\%28\%3F\%3A\%5E\%7C\%5C/\%29\%28\%5C.kiro\%5C/\%29/\&type=code}{42200} & - & - & - & 42.2K \\
\hline
\href{https://github.com/langchain-ai/open-swe}{Langchain Open SWE} & - & \heuristiclink{open-swe (412)}{https://github.com/search?q=Co-authored-by\%3A\%22open-swe\%22\&type=commits}{412} \heuristiclink{open-swe (author, 1.3K)}{https://github.com/search?q=author\%3A\%22open-swe\%22\&type=commits}{1300} & \heuristiclink{open-swe/ (1.2K)}{https://github.com/search?q=head\%3Aopen-swe/\%20type\%3Apr\&type=pullrequests}{1200} & \heuristiclink{open-swe (677)}{https://github.com/search?q=label\%3Aopen-swe\%20type\%3Apr\&type=pullrequests}{677} & 3.6K \\
\hline
\href{https://github.com/All-Hands-AI/OpenHands}{OpenHands} & \heuristiclink{.openhands/ (328)}{https://github.com/search?q=path\%3A/\%28\%3F\%3A\%5E\%7C\%5C/\%29\%28\%5C.openhands\%5C/\%29/\&type=code}{328} & \heuristiclink{openhands@all-hands.dev (35.9K)}{https://github.com/search?q=Co-authored-by\%3A\%22openhands\%40all-hands.dev\%22\&type=commits}{35900} \heuristiclink{openhands-agent (author, 34.5K)}{https://github.com/search?q=author\%3A\%22openhands-agent\%22\&type=commits}{34500} & \heuristiclink{openhands/ (440)}{https://github.com/search?q=head\%3Aopenhands/\%20type\%3Apr\&type=pullrequests}{440} & - & 71.1K \\
\hline
\href{https://opencode.site}{Opencode} & \heuristiclink{.opencode/ (1.6K)}{https://github.com/search?q=path\%3A/\%28\%3F\%3A\%5E\%7C\%5C/\%29\%28\%5C.opencode\%5C/\%29/\&type=code}{1600} & \heuristiclink{noreply@opencode.ai (5.7K)}{https://github.com/search?q=Co-authored-by\%3A\%22noreply\%40opencode.ai\%22\&type=commits}{5700} & - & - & 7.3K \\
\hline
\href{https://github.com/QwenLM/qwen-code}{Qwen Coder} & - & \heuristiclink{Qwen-Coder (4.6K)}{https://github.com/search?q=Co-authored-by\%3A\%22Qwen-Coder\%22\&type=commits}{4600} \heuristiclink{qwen-coder@alibabacloud.com (4.1K)}{https://github.com/search?q=Co-authored-by\%3A\%22qwen-coder\%40alibabacloud.com\%22\&type=commits}{4100} & - & - & 8.7K \\
\hline
\href{https://github.com/RooCodeInc/Roo-Code}{Roo Code} & \heuristiclink{.roo/ (6.7K)}{https://github.com/search?q=path\%3A/\%28\%3F\%3A\%5E\%7C\%5C/\%29\%28\%5C.roo\%5C/\%29/\&type=code}{6700} & \heuristiclink{roomote@roocode.com (3.4K)}{https://github.com/search?q=Co-authored-by\%3A\%22roomote\%40roocode.com\%22\&type=commits}{3400} \heuristiclink{roomote (author, 666)}{https://github.com/search?q=author\%3A\%22roomote\%22\&type=commits}{666} \heuristiclink{roomote (3.6K)}{https://github.com/search?q=Co-authored-by\%3A\%22roomote\%22\&type=commits}{3600} \heuristiclink{Roo Code (6.2K)}{https://github.com/search?q=Co-authored-by\%3A\%22Roo\%20Code\%22\&type=commits}{6200} & - & - & 20.6K \\
\hline
\href{https://www.sketch.com/}{Sketch} & - & \heuristiclink{hello@sketch.dev (1.7K)}{https://github.com/search?q=Co-authored-by\%3A\%22hello\%40sketch.dev\%22\&type=commits}{1700} & - & - & 1.7K \\
\hline
\href{https://sourcery.ai/}{Sourcery} & - & \heuristiclink{sourcery-ai (10.4K)}{https://github.com/search?q=Co-authored-by\%3A\%22sourcery-ai\%22\&type=commits}{10400} & - & - & 10.4K \\
\hline
\href{https://github.com/github/spec-kit}{SpecKit} & \heuristiclink{.specify/memory/constitution.md (1.6K)}{https://github.com/search?q=path\%3A/\%28\%3F\%3A\%5E\%7C\%5C/\%29\%28\%5C.specify\%5C/memory\%5C/constitution\%5C.md\%29\%24/\&type=code}{1600} & - & - & - & 1.6K \\
\hline
\href{https://sweep.dev/}{Sweep} & - & \heuristiclink{sweep-ai (1.0K)}{https://github.com/search?q=Co-authored-by\%3A\%22sweep-ai\%22\&type=commits}{1000} & \heuristiclink{sweep/ (28.0K)}{https://github.com/search?q=head\%3Asweep/\%20type\%3Apr\&type=pullrequests}{28000} & - & 29.0K \\
\hline
\href{https://github.com/eyaltoledano/claude-task-master}{Taskmaster} & \heuristiclink{.taskmaster/ (14.9K)}{https://github.com/search?q=path\%3A/\%28\%3F\%3A\%5E\%7C\%5C/\%29\%28\%5C.taskmaster\%5C/\%29/\&type=code}{14900} & - & - & - & 14.9K \\
\hline
\href{https://www.trae.ai/}{Trae} & \heuristiclink{.trae/ (4.4K)}{https://github.com/search?q=path\%3A/\%28\%3F\%3A\%5E\%7C\%5C/\%29\%28\%5C.trae\%5C/\%29/\&type=code}{4500} & - & - & - & 4.4K \\
\hline
\href{https://www.warp.dev/}{Warp} & \heuristiclink{WARP.md (3.2K)}{https://github.com/search?q=path\%3A/\%28\%3F\%3A\%5E\%7C\%5C/\%29\%28WARP\%5C.md\%29\%24/\&type=code}{3200} & - & - & - & 3.2K \\
\hline
\href{https://windsurf.com/}{Windsurf} & \heuristiclink{.windsurf/ (10.3K)}{https://github.com/search?q=path\%3A/\%28\%3F\%3A\%5E\%7C\%5C/\%29\%28\%5C.windsurf\%5C/\%29/\&type=code}{10300} \heuristiclink{.windsurfrules (2.3K)}{https://github.com/search?q=path\%3A/\%28\%3F\%3A\%5E\%7C\%5C/\%29\%28\%5C.windsurfrules\%29\%24/\&type=code}{2300} & - & - & - & 12.6K \\
\hline
\end{tabular}
\label{tab:heuristics}
\end{table*}

Table~\ref{tab:heuristics} shows the heuristics we have detected so far, across multiple categories of GitHub artifacts. Importantly, we do not expect all of these specific heuristics to remain unchanged in the long term (see \hyperref[per:velocity]{Peril 4 of velocity}). This is why we distill these heuristics in a set of more general detection strategies, across several categories of GitHub artifacts, that we expand on below.  



\setBoxStyle{margin}
\begin{Promise}{Presence of Traces}
Coding agents leave many visible traces in software repositories, in files, commits, issues, and pull requests, detectable by specific heuristics, and more generic detection strategies.\label{pro:traces}
\end{Promise}

\subsection{Files}

Several types of files indicate that a coding agent is used in a repository: \emph{Configuration files} toggle a variety of settings; \emph{Rules} and \emph{Guidance} files affect their behavior when working in the repository.

\paragraph{Configuration files} Like many tools, agents can be configured, and these settings are often found in configuration files that follow naming conventions. A repository containing such a configuration file indicates that an agent has been set up at least once in the repository. Moreover, depending on the specific agent, some of the settings inside the file can offer insights on how the agent works. 

In particular, some configuration files will contain information on the permissions the agent has been granted, \ie, which actions it can take autonomously (\eg reading files), which actions require developer approval (\eg writing files), and which actions are not allowed (\eg accessing the internet). Studies may use these settings to compare repositories where agents are granted high autonomy, versus repositories where agents are supervised more closely.

Other interesting settings may include the availability of MCP servers, that provide agents with additional tools. These settings provide valuable context on the usage of a given agent and its capabilities. Other settings can regulate how the agent advertises its presence via commits. In general though, the settings are highly tool specific (see the \hyperref[per:diversity]{Peril 3 of diversity}). Settings can be found either in a specific repository, but also, as some developers do, in repositories storing configuration specifically (dotfiles).

\emph{Rules and Guidance files} are more specific to coding agents. They contain natural language instructions that the agent should comply with when working in the repository. Almost always, these files follow the markdown format. These files are concatenated to the LLM's prompt to affect its behavior. The files might have a project-level scope (applying to the entire project), or a smaller scope (e.g., a module), in which case they are included in the prompt only if the agent is working in this module. These files may be human-written, but it is also common that they are generated by the agent. 

\emph{Rules} provide specific natural language instructions that the agent should follow. Example include following certain coding conventions, or how to perform specific actions. Some rules might be very generic, other very specific to the project (documenting a specific convention, running a command) \footnote{\href{https://github.com/adobe/data/blob/main/.cursorrules}{Click to browse a simple cursor rules file}}.

\emph{Guidance} files also primarily contain natural language, but tend to have a broader scope. They may document higher-level knowledge about a project or a module \footnote{\href{https://github.com/Windscribe/Android-App/blob/main/CLAUDE.md}{High-level project description example}}, such as its architecture and its main components, or provide rationale for certain decisions. They may contain detailed task descriptions highlighting the steps to solve a specific task \footnote{\href{https://github.com/marcusgoll/robinhood-algo-trading-bot/blob/c05474f63df1aa0a92aee061592ea25dfd13f6d9/specs/order-management/plan.md}{Example detailed implementation plan}}. In both of these cases, the files may be generated by the agent as summaries of the code base, or plans to implement a task.  Guidance files may also describe tactics and strategies that the LLM should use to tackle problems in general \footnote{\href{https://github.com/bigcapitalhq/bigcapital/blob/develop/.cursor/commands/speckit.clarify.md}{Example guidance with strategy}}. The space of instructions is very large, and is a promising area of study (see \hyperref[pro:study_pot]{Promise on Potential}).

Rules and guidance file encode \emph{tacit knowledge} about the project, or software development, or even problem solving in general. As such, they are valuable artifacts that are useful beyond a single developer. This is why they are often committed in the repository to be shared, although some developers elect not to commit them.

\paragraph{Detection} Many of these files follow established naming conventions, making them easy to detect (\eg, \texttt{CLAUDE.md}). In some cases, the files themselves may vary, but the directory they are in follows a naming convention (\eg, \texttt{.cursor/}). However, some naming patterns can have false positives (\eg, Aider guidance file is often called \texttt{CONVENTIONS.md}, but many repositories use this to document coding conventions for developers, not agents).

\subsection{Commits}
Several agents (but not all) can be given the capability to commit on behalf of the developer, and may (or may not) advertise it. Agents can either be given access to a specific tool to commit, or (more commonly), simply use their shell tool to commit on behalf of the developer. When authoring a commit, agents will often compose the commit message themselves, rather than asking the user, and summarize change in varying levels of details. They can advertise their presence in several ways:

\begin{itemize}
\item By adding themselves as a \emph{co-author} of the commit, with the agent's user advertised as the main author. In this case, they use the ``\texttt{Co-authored-by:}'' trailer, which is added to the commit. Of note, agents may not adhere to the convention exactly, occasionally using different casings.\footnote{\href{https://github.com/Num8398893/Num8398893/commit/23286b375c7706da993fd1a27ac8554a36d7253c}{Example co-authored commit}}
\item Less commonly, as the author of the commit, similarly to how bots such as dependabot operate.
\item Some agents also add additional trailers to the commit, such as ``\texttt{Generated by: Claude}''.
\end{itemize}
Depending on the agent or developer workflow, the commits may be authored by the agent completely autonomously, or include developer oversight or edits. 

\paragraph{Detection} Maintaining a list of known author/co-author names makes the detection of agent-authored commits straightforward. More advanced heuristics might be possible.

\subsection{Pull requests}

Some agents workflows include Pull Requests (PR): the agent will author one or more commits, and additionally author a PR that is submitted for review on GitHub, with an appropriate summary. A PR is richer than a commit;
notably, it allows other GitHub users to comment on the PR, and for developers to decide to merge the PR or not. Some agents may revise their work when a developer comments on the PR and indicates that additional changes should be made, often also appending a comment in response (in addition to a commit).
Bots submitting PRs are not new (\eg dependabot), but the PRs submitted by agents can have a much broader scope. Depending on user instructions and the capabilities of the agents, submitted PRs can range from simple bug fixes to entire new features.\footnote{\href{https://github.com/microsoft/vscode/pull/271364}{Example PR with user/agent interaction, implementing a feature}}


\paragraph{Detection} Some agents follow patterns when working with PRs, notably by advertising their presence in the name of the branch used in the repository (\eg \texttt{codex/branch-name}). Some agents also assign labels to their PR. In addition, agents may submit their PR either as the user that triggered the agent, or as a specific user that can be detected. Finally, agents may interact with developers in the PR itself via comments.

\subsection{Issues}
Issues submitted by and assigned to bots are not new. Some agents support the same kind of interactions, especially when they act as a specific user or are associated with GitHub actions. The issue in this case acts as a task description for the agent. This is interesting since it gives additional visibility on the process. As documented in \hyperref[per:partial_obs]{the Peril of Partial Observability}, the initial prompt given to the agent is not necessarily documented; issues solve this in some case.\footnote{\href{https://github.com/OpenHands/docs/issues/36}{Example with a user assigning an issue to OpenHands, which creates a PR in response}}

\paragraph{Detection} Participation in issues can be detected if the user associated with the agent is mentioned or responds in the issue.

\subsection{Users}
Like other bots, some agents are associated with specific GitHub users. Although users may not carry much information, their presence is a signal in itself. In addition, they allow some of the other functionality, such as assigning an issue to an agent, or allowing the agent to comment on pull requests.

\paragraph{Detection} Via a list of known agent users. 

\section{Promises, perils and mitigations} 
\label{sec:promises-perils}
\subsection{Coding agent adoption is real}

\begin{Promise}{Adoption is real and significant}
As of October 2025, the adoption of coding agents is already clearly visible on GitHub.\label{pro:adoption}
\end{Promise}

As of mid-October 2025, our study of the extent of adoption of coding agents on GitHub estimates that between 15 to 19\% of GitHub projects show traces of coding agents. Furthermore, the adoption is growing steeply. Thus, we expect future adoption to grow significantly in the coming months and years, and for coding agents to take a larger place in software development.

In addition, our study identified repositories with a wide variety of coding agent use, from the experimental (a handful of commits co-authored by a coding agent), to the pervasive (repositories where the \emph{majority} of commits are co-authored by coding agents). 

To give additional insight on the wealth of information available, \tabref{tab:heuristics} provides some counts on the frequency of each heuristics via simple queries to GitHub's web interface in order to estimate their count. While this count is not precise, does not offer us information about the number of repositories in which the heuristics are found (\ie, some heuristics may match with a repository multiple times), and may contain false positives, we can already get a sense of the adoption. Notably, the most popular tools have been visibly used to author \emph{millions} of commits and pull requests.


\begin{Promise}{Study potential}
The visible traces left by agents give us a window in their impact on software engineering.\label{pro:study_pot}
\end{Promise}

For the first time, agent-based automation is visible, \emph{at scale}, thanks to the traces generated by coding agents. Previous studies using MSR techniques, such as the one of Tufano \etal~\cite{tufano2024unveiling} or the one of Xiao \etal~\cite{xiao2025self_admitted_ai}, relied on deliberate annotations by developers; while very valuable, these studies could not have the same level of exhaustiveness. 
Coding agents traces allow us to finally reliably detect LLM-generated code, unveiling a level of automation that was previously hidden. Given the wealth of traces available, we think that many MSR studies could be done on the adoption and impact of coding agents in software engineering.

For instance, while the GitClear whitepapers~\cite{hardinggitclear2024, hardinggitclear2025}, in the absence of stronger signals, used a time-based delimiter (``before Copilot'' vs ``after Copilot''), mining coding agent traces allow us to compare repositories that explicitly use agents with those that do not. The traces are also much more precise, as they have fine degree of granularity, down to individual commits and PRs. 

Studying code produced with coding agents will allow us to have a clearer understanding of the impact of coding agents on issues such as code quality, or code defects. Importantly, longitudinal analyses of repositories can study the longer-term impact of coding agent use. Analyzing commits and PRs over time will give us insights on the true impact of coding agents on developer productivity. Analyzing the success of PRs (in terms of merge rates, rework, \etc), can give us insights on the factors that influence the success or failure of coding agents. These may be broader than individual PRs: analyses of characteristics of repositories (\eg how does a repository's code quality affect agent performance?) and their impact on the use and success of coding agents will be insightful. For instance, we expected to find higher adoption of coding agents in smaller repositories; we were surprised to find that larger repositories had comparable (if not higher) rates of adoption \cite{agentadoptionpaper}.



\begin{Promise}{Mining developer--agent interactions}
The visible traces left by agents offer a unique opportunity to observe how \emph{developers} interact with AI code models.\label{pro:dev_agent_int}
\end{Promise}


While interactions between developers and LLMs were previously confined to private chat environments, the emergence of coding agents as active participants in shared social spaces (\eg pull requests, issues, code reviews) presents new opportunities to systematically study human--AI collaboration. The interaction traces left by these agents capture not only their outputs but also developers’ responses:~accepting, rejecting, or refining their contributions. Unlike conventional LLM use, these public traces reveal how developers articulate their needs, review and correct the agent contributions, and steer their behavior.
They can also reveal how developers might restructure their workflows and documentation to ``make space'' for agents to effectively contribute to their projects.

Mining these interactions can enhance our understanding of the task distribution between developers and agents, identifying which activities developers delegate to agents and which they retain under their control. It also enables analysis of review dynamics, including how developers critique or modify agent outputs and, conversely, how agents propose changes to human-written code. More broadly, these records provide a unique perspective on the successes and failures of coding agents in practice, exposing frictions and pain points in developer--agent interactions. This information will inform the design of new agents and allow for a better integration of their capabilities into developer spaces and workflows.

\begin{Promise}{Study agent knowledge}
The files that coding agents leverage in their tasks are a very rich source of information.\label{pro:study_agent_knowledge}
\end{Promise}

Agent rules and guidance files are critical for coding agents to work well. As mentioned in \secref{sec:agents}, these files can contain several kinds of valuable knowledge: 1) rules and conventions that agents should follow; 2) more general knowledge about the repository (e.g., high-level organization); 3) detailed task descriptions and plans; 4) general tactics and strategies to solve problems. While this is an early classification, we believe that a full qualitative study of the contents of these files will be extremely instructive and valuable to know more about how agents are used in practice. For instance, while most repositories have straightforward guidance files, some users go to considerable lengths to define extensive and sophisticated guidance, covering a variety of dimensions (\eg documenting project context, technical choices, architectural patterns, \etc). 

After having built a better understanding of the knowledge in these files, many other studies are possible. Possible studies include for instance studies of the effectiveness of the rules encoded in the files (\eg, to which extent do coding agents follow the conventions?), or their impact on the success of coding agents (\eg, to which extent do high-quality specifications influence coding agent success?).

Finally, agent configuration could also allow studies, \eg whether and how the availability of advanced tools via the Model Context Protocol or similar mechanisms is leveraged by agents.
%

\subsection{Agent detection: the devil is in the details}

\begin{Peril}{Partial observability}
We can not observe all agent activity, only parts of it.\label{per:partial_obs}
\end{Peril}
\noindent
Mitigations: 
(1) \hyperref[mitig:more_heuristics]{Additional heuristics} (2) \hyperref[mitig:more_data]{Gather more data}
\smallskip

While coding agent leave valuable traces, they offer only a partial view of the activity of coding agents. Moreover, due to the \hyperref[per:diversity]{Peril 3 of diversity}, some of the partial information is agent-specific. The missing information can come from several dimensions.

First, different agents or developer workflows will share different types of information. For instance, while several agents sign co-authored commits, others do not. Some agents may not support committing on behalf of the developers and, even when they do, developers may not be willing to adopt this workflow, and may instead commit manually. Finally, even if a developer does use an agent that commits on their behalf, they may not be willing to advertise it. Some agent configuration files (\eg Claude Code) will co-sign commits by default, but have some settings to disable this feature; alternatively, we have seen examples of guidance files explicitly telling the agent \emph{not} to sign commits under any circumstances. As a consequence, our study of coding agent adoption, finds that more than 40\% of projects that have markers of usage of coding agents do not have any markers at the commit level \cite{agentadoptionpaper}, making it difficult to ascertain to which degree these projects have adopted coding agents (they may, after all, simply not be using the agent).

The converse is true: due to differences in terms of workflow, agents may advertise their presence via commits and pull requests, but not via files. Increasingly, other agents may use standard files such as \texttt{AGENTS.md}, making the detection of specific agents more difficult. Finally, even if agent rules and guidance files are useful knowledge that should be shared in the repository, a significant portion of are not. Our coding agent adoption study estimates that 20\% of projects that have agent guidance or configuration files choose to exclude all of them from commits via the \texttt{.gitignore}. Harder to quantify, developers may store their agent configuration in separated repositories (\eg a \texttt{dotfiles} repository).

Even when commits, pull requests and files are present, they only represent the final output of the coding agent. In general, what was the input (the initial prompt) is not present, except when the agent is explicitly assigned an issue (in which case the issue description is the prompt). Depending on their guidance, agents may be asked to write down plans before implementing changes, which is a step removed from the initial prompt, but is certainly closer.

Also missing in most cases is the amount of involvement of developers in supervising the agent. Since only the end result is shown in commits, it is hard to know if the developers closely supervised the agent (or edited the code) or not. Here, pull requests may be more informative, particularly if they contain developer-agent interactions via comments, showing part of the interactions.

Finally, several agents (\eg Codex) do record very detailed traces of the agent's execution, including prompts, interactions, and tool calls (much like is available on SWE-bench~\cite{bouzenia2025understanding}). In the case of Codex, links to these traces is even provided as part of the pull request. However, these links require authentication on OpenAI's servers, and are not publicly accessible.

\mitigation{Additional heuristics}{mitig:more_heuristics}
Additional heuristics can remediate parts of the \hyperref[per:partial_obs]{Peril 1 of Partial observability}. The first is to identify repositories where it is very likely to apply, by identifying know patterns that reduce observability. Examples include the presence of \texttt{.gitignore} files excluding agent configuration files, checking for known settings that affect the agent's harness (\eg disabling commit signing), or guidance with similar goals (although this is more challenging to detect).

A second way is to detect artifacts that are likely to be made at least partially by coding agents. Several heuristics can be devised for this, exploiting for instance LLM tendencies to thoroughly document their work in commit messages or pull requests (often better than developers), or their usage of Emojis. Other heuristics might use process metrics (\eg looking for changes in the rate of commits or pull requests~\cite{li2025rise}). Work to do this already exist for text snippets~\cite{mitchell2023detectgpt}, or for code~\cite{suh2025empirical}, but repository-level heuristics and information might prove useful signals. Needless to say, these heuristics should be evaluated carefully. In this regard, the availability of data with clear markers for usage of coding agents might be useful to define training, validation, or test sets to evaluate simple or elaborate heuristics to detect unlabeled coding agent activity. On the other hand, changes to model capabilities over time may change the nature of the problem, making detection an arms race between improving models and detectors.

\mitigation{Gather more data}{mitig:more_data}
Another mitigation is to collect more complete data. Similarly to how developer interaction data can be collected via telemetry~\cite{maalej2013collecting}, some agents provide functionality to record more information, via explicit telemetry~\cite{openai_codex_config_observability} or user-defined hooks~\cite{cursor_agent_hooks, anthropic_claude_code_hooks}. This is however a significant undertaking.

\begin{Peril}{Agent multiplicity}
There are many agents, leaving traces in very different ways. \label{per:multiplicity}
\end{Peril}
\noindent
Mitigations: 
(1) \hyperref[mitig:unequal_adoption]{Unequal adoption},
(2) \hyperref[mitig:knowledge_sharing]{Knowledge sharing}. \smallskip

As can be seen in \autoref{tab:heuristics}, there are large number of agents, with a wide range of different heuristics. This number is still growing: while the first were released in 2024, and there were a dozen in early 2025, the last few months have seen the total reach three dozen agents. The large number of agents make it more challenging to keep track of them, and to search for effective heuristics in all cases. Using only some heuristics will miss agent uses, which may be problematic for a study that values exhaustiveness.



\basemitigation{Mitigation for Perils 2 and 3: Unequal adoption}{mitig:unequal_adoption} \textcolor{red}{Move after Peril 3?}
While many different agents are available, most of the adoption is focused on a few agents.
Currently, Claude, Codex, Cursor and Copilot are the most popular and captures more than 80\% of the adoption, whereas agents like OpenHands or Jules see comparatively little usage. This can be seen in \tabref{tab:heuristics} by looking at the counts of each heuristic; our study of agent adoption points towards a similar conclusion. Thus, studies can focus on a subset of agents and capture a majority of the usage. That said, studies of smaller agents are still useful, to study their specific characteristics and the impact of these characteristics on the way they work.


Previous work on JavaScript front-ends frameworks~\cite{ferreira2022adoption} or version control~\cite{orehovavcki2020modelling} show that tool adoption is highly dependent on popularity of the framework, creating a phenomenon of ``the rich get richer''.
While this is still quite early to observe this effect, and this might change in the future, we see the emergence of a winner takes all phenomenon, in which a very small subset of coding agents captures most use, leading to less agents in the long term.

\begin{Peril}{Diversity}
Each agent works differently, and this may be necessary to consider in any data analysis.\label{per:diversity}
\end{Peril}

\noindent
Mitigations: 
    (1) \hyperref[mitig:unequal_adoption]{Unequal adoption},
    (2) \hyperref[mitig:knowledge_sharing]{Knowledge sharing}.
\smallskip

There is a wide range of ways agents work. Each agent has a different harness, which may expose different capabilities. For instance, some agents such as Codex submit their changes via pull requests; others (\eg Claude) tend to use commits using the Co-authored-by trailer, or other conventions (\eg full authorship). 

Beyond the need to develop specific heuristics, these changes affect the workflow developers follow while using the tools, which, depending on the study, may be important to take into account. Examples of this include the way that agents work with pull requests: PRArena~\cite{prarena2025} is a website that tracks, for a set of agents, the merge rate of their pull requests, using heuristics similar to the ones presented in this paper. 
Some agents, such as Codex, favor private iteration on pull requests, while others, such as Copilot, will submit a pull request and favor public iteration with developer via GitHub pull request comments. This is clearly visible in the difference between the PR merge rates: among all pull requests, Codex has a much higher merge rate (86.6\% on 20/10/2025) than Copilot (63.2\%); however, if one excludes ``draft'' pull requests, Copilot's merge rate jumps to 93.1\%. Even more importantly, PRArena has no information on Claude Code, despite it being one of the most prominent agents (our study finds it has the highest adoption level): this is because Claude Code's workflow did not include pull requests until very recently (september 29, 2025). The problem is broader: while some agents can author pull requests or commits, other agents do not (also see the \hyperref[per:partial_obs]{Peril 1 of Partial observability}).

Another dimension of variability is how agents use guidance files. Earlier agents such as Cursor put more emphasis on simpler rules (possibly reflecting the limited capacities of earlier models). Most agents now use free-form guidance files (\eg AGENTS.md, CLAUDE.md), where the user (or the agent) can summarize information deemed important. Some agents incentivize their users to adopt sophisticated guidance systems (\eg Cline's memory bank pattern, or Kiro's steering). As these instructions vary considerably, we expect them to influence the behavior of the agents.


This is mitigated by the previous two mitigations: only a few agents are widely used and this will keep going with less agents in the long term making it easier for data analysis.


\begin{Peril}{High velocity}
The state of the practice evolves very quickly, and this has far reaching implications.\label{per:velocity}
\end{Peril}

\noindent
Mitigations: 
(1) \hyperref[mitig:revisit]{Revisit studies}, (2) \hyperref[mitig:standardization]{Standardization}, (3) \hyperref[mitig:knowledge_sharing]{Knowledge sharing}.
\smallskip

The adoption of agents, their functionality, and their usage changes rapidly, for many reasons. First, the capacities of the underlying LLMs is still evolving, and this influence the agents capabilities. Kwa \etal estimate that the duration of tasks that LLMs can solve autonomously doubles every 7 months~\cite{kwa2025measuring, kwa2025measuring2}. This impacts the tasks that are delegated to coding agents over time. 

This rate of change is also visible in the \hyperref[per:multiplicity]{Peril 2 of Agent multiplicity}: many agents exist; new ones are regularly introduced; and existing ones change, which sometimes impact the heuristics. For instance, Cursor transitioned from storing rules in a single \texttt{.cursorrules} file to a \texttt{.cursor} directory for storing more general guidance. Another example is the \texttt{AGENTS.md} file: originally, this was the default file used for guidance by the Codex agent. There is now a standardization effort around this format, with a growing number of coding agents supporting it. As a consequence, this heuristic is no longer a marker for the Codex agent, but rather a general marker for coding agent usage in a given repository. In addition, since guidance files are text files, it is easy for an agent to use the guidance of other agents: for instance, Zed's agent looks for files used by other popular agents and incorporates it as its own guidance~\cite{zeddev2025rules}. 

As a consequence, the detection of agent traces needs to be regularly updated to keep up with the fast pace of releases. Combined with the perils of \hyperref[per:multiplicty]{multiplicity} and \hyperref[per:diversity]{diversity}, this is challenging, which is why we introduce heuristic repository (see \secref{sec:repository}). If the trend of standardization continues, it is possible that it will become both \emph{easier} to detect agent use in general, but \emph{more difficult} to detect the usage of individual agents; this will make it more challenging to take into account the diversity of coding agents.

Another manifestation of the \hyperref[per:velocity]{Peril 4 of velocity} is the importance of data freshness. While datasets are very important in MSR research, changes to coding agents mean that datasets of their activity will age more rapidly than traditional MSR datasets, particularly if the capacities of coding agents continue to evolve rapidly.

\mitigation{Revisit studies}{mitig:revisit}
Given the changes to the capacities of LLMs and their accompanying changes to the capabilities of agents, and the use that is made of these agents, we think that some studies may benefit from being revisited regularly in order to check how their conclusions evolve with time. For instance, a study on the quality of code generated by agents may be revisited if one suspects the evolving capacities may have resulted in higher code quality. Automation helps; studies that use manual annotation will be more difficult to revisit, although partial automation with LLMs can help \cite{ahmed2025llm}.

\mitigation{Standardization}{mitig:standardization}
There are initiatives such as the website AGENTS.md in order to build standards, as it is the case for the MCP protocol. Some projects adopt guidelines to recognize and standardize contributions by coding agents. Perhaps in the near future more standards will emerge, simplifying the studies.





\subsection{Beyond agent detection}

\begin{Promise}{Model upgrades}
Coding agent traces enable the study of the impact of new model release on the use of coding agents.\label{pro:model_upgrade}
\end{Promise}

As the capabilities of the underlying models drive the capabilities of agents and their usage, studies of their actual impact would be very useful. Since models have clear release dates, and since some coding agents have large amounts of data, this becomes feasible, for instance by comparing commits done in time periods in which an updated version of a model was made available, with a time period where an older version was available.




\begin{Peril}{Multiple Models}
Some coding agents support multiple models, so determining which model is used is not trivial.\label{per:model_agents}
\end{Peril}
Several agents provide multiple models for their users, yet the traces left do not specify which model was used.  For instance, Copilot can be used with GPT-5, Sonnet and many other, yet there is no indication of the model used.
This limits the ability to study the impact of a specific model release. Coding agents specific to a model vendor are much more likely to use a model from that vendor, however, and some coding agents (such as Aider) do specify which model was used to author a commit. In other cases, the agent's configuration files may provide some clues, as the models to use can be specified there.




\begin{Peril}{Scientific reproducibility}
Coding agents are non-deterministic and often based on closed, proprietary LLMs, undermining scientific reproducibility.
\label{per:repro}
\end{Peril}

\noindent
Mitigations: 
(1) \hyperref[mitig:open_llm]{Open coding agents},
(2) \hyperref[mitig:replic]{Replication packages}.
\smallskip

LLMs are non-deterministic~\cite{bhaskar2024reproscreener}: they can provide different results starting from identical inputs (project status, prompt, etc.), and even with a fixed random seed (with batch inference \cite{he2025defeating}). Furthermore, most coding agents are currently based on LLMs that are closed and/or proprietary~\cite{liesenfeld2024osaiindex}, as a whole or in part~\cite{white2024mof}, and accessed via remote APIs.
In other cases the models are available, but only provide the final model weights. 
In rare cases (e.g., OLMo~\cite{olmo2}), all LLM parts---training datasets, training pipeline, model weights, inference code---are available and released under traditional open source licenses~\cite{liesenfeld2024osaiindex, white2024mof}. The agent themselves, implemented in software, are available under licenses with varying degrees of openness, while the inference stack is most often opaque and can affect performance~\cite{anthropic2025postmortem}. The inference stack may include ``model routers'' that allow to switch the underlying model, based on the query.

These factors make it either impossible or very hard to replicate the activities of coding agents, potentially undermining the scientific reproducibility of empirical experiments that analyze them. 

MSR research will focus on existing traces, which overwhelmingly come from ``closed'' systems.
For studies that focus on traces only, this might not be as problematic, but specific study designs may face more limitations.
Studies that require running coding agents (\eg a study of agent success/failure factors that whishes to establish a causality link) or LLMs (\eg, to automate manual annotations) will face reproducibility issues. Closed models or inference stacks make it impossible to archive the models \footnote{API models access can also be deprecated and suspended, such as OpenAI's original Codex model.} or even precisely identify the used versions, as they can be silently altered by the model operators, compounding the Peril of Multiple Models.
Likewise, research interested in the impact of training data will not be possible with such closed systems.




\mitigation{Open coding agents}{mitig:open_llm}
Studies can trade off data volume for increased reproducibility by focusing on the most open coding agents. Using established scales such as the Modeling Openness Framework~\cite{white2024mof} or the European Open Source AI Index~\cite{liesenfeld2024osaiindex} will help.
Note that, even with fully-open agents, the problem of LLM non-determinism remains.

\mitigation{Thorough replication packages}{mitig:replic}
Empirical studies conducted using coding agents should report agent information and/or artifacts in replication packages.
For open agents, the package should include all relevant artifacts (datasets, models, agent code, etc.).
For closed agents, the package should document as best as possible their versions and deployment details.

\begin{Peril}{Costs shape usage}
Coding agents are expensive, and this influences their usage.\label{per:cost}
\end{Peril}
\noindent
Mitigation:
(1) \hyperref[mitig:strat_sampling]{Strategic Sampling} 
\smallskip

LLM inference is compute intensive, particularly for coding agents.
For a single inference turn, processing the LLM's prompt has a quadratic complexity with respect to the prompt size, while generating tokens has a linear complexity (with respect to the prompt and the previously generated tokens) due thanks to the key-value cache.
This makes inference on long prompts much more expensive than shorter prompts.
Coding agents tend to work on tasks requiring large prompts, including guidance, relevant context, tool results, built over multiple turns (\eg OpenHands routinely consumes more than a million tokens on SWE-bench tasks~\cite{bouzenia2025understanding}). 

As a consequence, using coding agent is expensive. The 2025 Stack Overflow survey~\cite{stackoverflow2025ai} shows that the majority of interviewees agree that ``the cost of using AI agent platforms is a problem''. Different pricing models exist: flat-rate (with usage limits) and usage-based. This introduces a discrepancy between users depending on their resources: open-source contributors might have to buy their own subscription, opting for lower tiers than developers working for companies that pay for their access. A few testimonials advocate that budgets exceeding 1,000\$ per developer per month can be justified~\cite{quigley2025claude}.  This, in turn, can lead to radically different usages between open-source repositories and industrial repositories, where open-source repositories and contributors, vastly more prevalent on GitHub, may feature lower usage of coding agents. The effect of this is broader, with developers from different countries being more or less price-sensitive. In addition price-sensitive users may also use multiple accounts with free tiers to satisfy their usage demands. Conclusions on the usage of coding agents should be made with this context in mind.
This also implies that the distribution of repositories using coding agents may be radically different than the distribution of open source repositories.

Furthermore, as mentioned in the \hyperref[per:velocity]{Peril of High Velocity}, the practice changes very quickly, including pricing: coding agents change their prices or usage limits over time (\eg Cursor introducing rate limits~\cite{truell2025cursorpricing}), model inference cost change, which will affect usage.


%


\mitigation{Strategic Sampling}{mitig:strat_sampling}
While there are some usage differences between users that have higher resources than others (especially those coming from industry), looking for open-source repositories in which companies contribute significantly may be helpful in finding users with usages more representative of those found in industry.

\begin{Peril}{AI coding slop}
Coding agents may produce more and larger development artifacts, of lower quality than those produced by humans. This ``AI coding slop'' will make the work of researchers in empirical software engineering harder.\label{per:coding_slop}
\end{Peril}

Traditionally, software development artifacts have a high signal-to-noise ratio and require significant human effort to produce.
For example, it can take a full day for a developer to fix a thorny bug; at the end of the day the observable artifact might be a one-line patch, recorded in Git with an accompanying detailed commit message.
Net result: significant human effort, small artifact of high quality.

\emph{``AI slop''}~\cite{hern2024aislop} is a derogatory term that refers to AI-generated digital content characterized by its abundance, low quality, and the perceived little effort put into generating it (causing its low quality).
Coding agents risk polluting the primary sources that researchers in empirical software engineering study (VCS repositories, collaborative coding platforms, etc.) with \emph{``AI coding slop''}: abundant, verbose, and poor-quality development artifacts (commits, pull/merge requests, comments, etc.). Examples include AI-generated security reports \cite{haxx2025death}, although discerning AI use is much more useful \cite{haxx2025newbreed}.

Although it is too soon to understand the magnitude of this problem, if AI coding slop is coming, it will materialize at different levels.
First, the number of artifacts researchers will have to analyze will increase significantly.
For example, in our experience it is already very common to observe very long interaction sequences among developers and agents in pull request discussions, publicly displaying the feedback loop needed to converge to an acceptable result.
Previously, the feedback loop was in the head of (or conversations among) developers and did not pollute interactions that researchers study.
Second, the size of individual artifacts will also increase, due to a natural tendency of LLMs to be more verbose than humans.

AI slop spreading to software development artifacts, will require methodological adaptations.
Studies only interested in analyzing human interactions will now need to filter out agent-human and agent-agent interactions.
This is not an entirely new requirement—MSR researchers already have to deal with bot detection—but will require new techniques, tools, and heightened methodological scrutiny.

\section{Heuristics and data repository}
\label{sec:repository}

\basemitigation{Mitigation for Perils 2, 3, and 4: Knowledge sharing}{mitig:knowledge_sharing}
The heuristics we have presented are necessarily numerous, varied and change prone due to the Perils of \hyperref[per:multiplicity]{Multiplicity}, \hyperref[per:diversity]{Diversity}, and \hyperref[per:velocity]{Velocity}. This makes keeping them up to date challenging. We think that the best mitigation of this is to distribute the effort, and share the knowledge. To this extent, we have started a repository to easily share this knowledge with the MSR community, and update it over time. The repository~\cite{agentminingrepo} is in its early stages, but it contains:

\begin{itemize}
\item A brief description of each coding agent (to be expanded over time), with links for further information;
\item A list of heuristics to detect various markers of its presence in software repositories with their period of validity.
\item For heuristics that allow it, links to GitHub queries to allow interactive browsing;
\item For heuristics that allow it, python scripts to gather sample data automatically; 
\item A list of repositories (ca. 10,000) featuring agent adoption as of October 2025, allowing targeted data collection. 
\end{itemize}

We call on the MSR community to contribute to this effort, by expanding the heuristic repository with additional heuristics as the practices changes, and by giving feedback on the effectiveness of the existing heuristics, in order to continuously improve them. While a handful of researchers will have difficulties keeping up with the velocity of coding agents, we hope that the much larger MSR research community, will do so in a much more scalable way.

\section{Discussion}
\label{sec:discussion}

\paragraph{Limitations}
This paper is focused on the promises and perils of mining coding agent usage, rather than the promises and perils of coding agent themselves. Issues such as their environmental impact, impact on the workforce, or intellectual property, while extremely important, are thus not covered in this paper. Previous work on promises and perils of mining Git~\cite{bird2009promises} and GitHub~\cite{kalliamvakou2014promises} also apply.

Our promises, perils, and heuristics are based on our personal experience conducting MSR studies of coding agents. While we strive for exhaustivity, we can not guarantee it. In fact, the \hyperref[per:velocity]{Peril 4 of velocity} implies that parts of the knowledge of this work (at least the heuristics) will need to be updated over time. The heuristic and data repository is a key mitigation to ensure this work remains current; we call on the community to contribute to this effort. 

By definition, heuristics are noisy. We take this in consideration when evaluating individual heuristics. For instance, this led us to exclude a heuristic for Aider, a guidance file named \texttt{CONVENTIONS.md}, as most such files contains content aimed at developers, rather than coding agents. Depending on the degree of accuracy sought by a study, it is possible for some heuristics to need additional filtering (such as date ranges) to refine the data. In addition, several heuristics might identify the same artifact. We call on the MSR community to document these cases. 

\paragraph{Implications}

The principal implication of this work is that the existence of coding agent traces opens up a vast field of potential studies of this phenomenon and its impact on Software Engineering. We provided some examples of such studies, but expect many more.

The implications in terms of potential studies vary with the degree of effective adoption of coding agents. The examples we provide (\eg, studies of effects on productivity, or of the factors influencing the success of agents) assume a ``business as usual'' scenario, were developers use coding agents to assist them in the same tasks that they work on today. However, some have much more ambitious visions for coding agents. If the observed trend of doubling the autonomy of agents on tasks every seven months is sustained for a few years~\cite{kwa2025measuring2}, the capabilities of coding agents will improve significantly, changing the tasks they are given. At the same time, techniques and tools to handle multiple agents in parallel emerge~\cite{devswarm_ai, warp_agents}, which would maximize this effect. If these trends materialize, then the type, the amount, and the scope of tasks delegated to agents may change rapidly, in which case software engineering practices themselves will be affected. 

We stress that the impact for MSR studies is potentially broader than this: the already high level of adoption shows that agents are already used in practice extensively. As mentioned in the Peril of AI Slop, if, as we expect it to, this trend continues, research that aim at studying \emph{human} contributions will need to take this into account, and \emph{exclude} contributions from agents, much like bot activity should be excluded. Our heuristics provide a first step for this. Moreover, if coding agents become very highly adopted, exactly what is a human contribution is an open question: if everyone uses coding agents, is everything ``slop''?

\section{Conclusion}
\label{sec:conclusion}
In the span of a few months, coding agents have transitioned from an academic endeavor to products that are used daily by developers. While LLM-based completion was used to streamline coding activities and was difficult to observe using MSR techniques, coding agents can address more tasks, and leave explicit traces in software repositories, finally enabling the study of AI-assisted coding via MSR techniques at scale. In this work, we presented heuristics to detect coding agent activities in files, commits, issues, and pull requests. More importantly, after having used these heuristics in MSR studies, we have distilled the promises and perils of mining coding agent activities. The principal perils being the \hyperref[per:multiplicity]{multiplicity} of \hyperref[per:diversity]{diverse}, \hyperref[per:velocity]{fast changing agents}, we invite the MSR community to contribute to our repository of shared knowledge and heuristics. The \hyperref[pro:study_pot]{main promise being the very large potential for studies of this phenomenon} and its impact on SE practices, we hope the community will join us in exploring it.
\clearpage

\bibliographystyle{ACM-Reference-Format}
\bibliography{main,agents}

\end{document}